\documentstyle[epsf,sprocl]{article}

\def\half{\frac{1}{2}}

\def\etal{{\it et~al}}
\begin{document}

\title{CLUSTER MASS PROFILE FROM LENSING}

\author{Hanadi AbdelSalam, Prasenjit Saha}

\address{Astrophysics, Department of Physics, Keble Rd., OX1 3RH, 
Oxford, UK}  

\author{Liliya L.R. Williams}

\address{Institute of Astronomy, Madingley Rd., CB3 0HA, Cambridge, 
UK}

\maketitle

\abstracts{ 
We propose a new technique to reconstruct non-parametrically
the projected mass distribution of galaxy clusters from their
gravitational lens effect on background galaxies. The beauty of our
technique, is that it combines information from multiple imaging
(strong lensing) and shear (weak lensing) as linear constraints on the
projected mass distribution. Moreover, our technique overcomes all the
drawbacks of non-linear methods. The method is applied to the first
cluster-lens A370 and to the cluster-lens A2218 which is exceptionally
rich in multiple images and arc(let)s. The reconstructed maps for both
cases revealed unprecedented level of details.  
}

\section{Introduction}

Gravitational lensing provides us with the necessary information
(constraints) needed for probing the matter distribution of cosmic
lenses. We present a new technique that combines information from
strong and weak lensing in a linear fashion and reconstructs mass maps
that follows the cluster light distribution as closely as possible
while strictly preserving the lensing constraints. Our technique
overcomes many of the drawbacks of previous methods: (a) It's
non-parametric, like Kaiser-Squires and related weak lensing methods
but unlike previous strong lensing work, (b) Error estimates are easy,
(c) Lensing constraints remains linear in all regimes (strong, weak \&
intermediate), (d) No mass sheet degeneracy and no end (boundary)
effects. 

\section{Outline of the method}

The technique works with a pixellated mass distribution, i.e
the ${\rm mn}$-th pixel is a square tile\footnote{Other
types of pixels are also viable and produce similar mass maps.} with a
surface mass density $\kappa_{\rm mn}$ in units of the critical
density. For a source at an unlensed position $\beta$, the scaled time
delay in a direction $\theta$ is
\begin{equation}
  \tau(\theta)=\half(\theta-\beta)^2-\frac{D_{\rm ls}}{D_{\rm s}}
	\sum_{\rm mn}\kappa_{\rm mn}\psi_{\rm mn}(\theta)\;, 
\label{eq:1}
\end{equation}
where $\psi_{\rm mn}(\theta)$ (a known function) is how much the ${\rm
mn}$-th pixel would contribute to the potential. 

Cluster lensing observations provide us with the following
constraints: \newline
(i)-- Image positions, say $\theta_1$, implying
\begin{equation}
\nabla\tau(\theta_1)=0.
\end{equation}
(ii)-- Shear measurements at some $\theta_1$; say the shear is
observed to be at least $k$, and aligned along the $\theta_{x^\prime}$
direction. This implies
\begin{equation}
k\left|\frac{\partial^2}{\partial\theta_{x^\prime}^{2}}\tau(\theta_1)\right|
\le
\left|\frac{\partial^2}{\partial\theta_{y^\prime}^{2}}\tau(\theta_1)\right|.
\end{equation}
Both types provide linear constraints on the unknowns $\beta$ and
$\kappa_{\rm mn}$.

By quadratic programming, we can produce the mass maps that e.g.,
minimize $M/L$ variation {\it while obeying the lensing
constraints}. By minimizing $M/L$ variations with respect to
Monte-Carlo light distributions, we can estimate errors. 
\section{Abell 370}
\begin{figure}[htb]
\parbox[c]{0.42\textwidth}{%
\epsfxsize=\hsize\epsffile{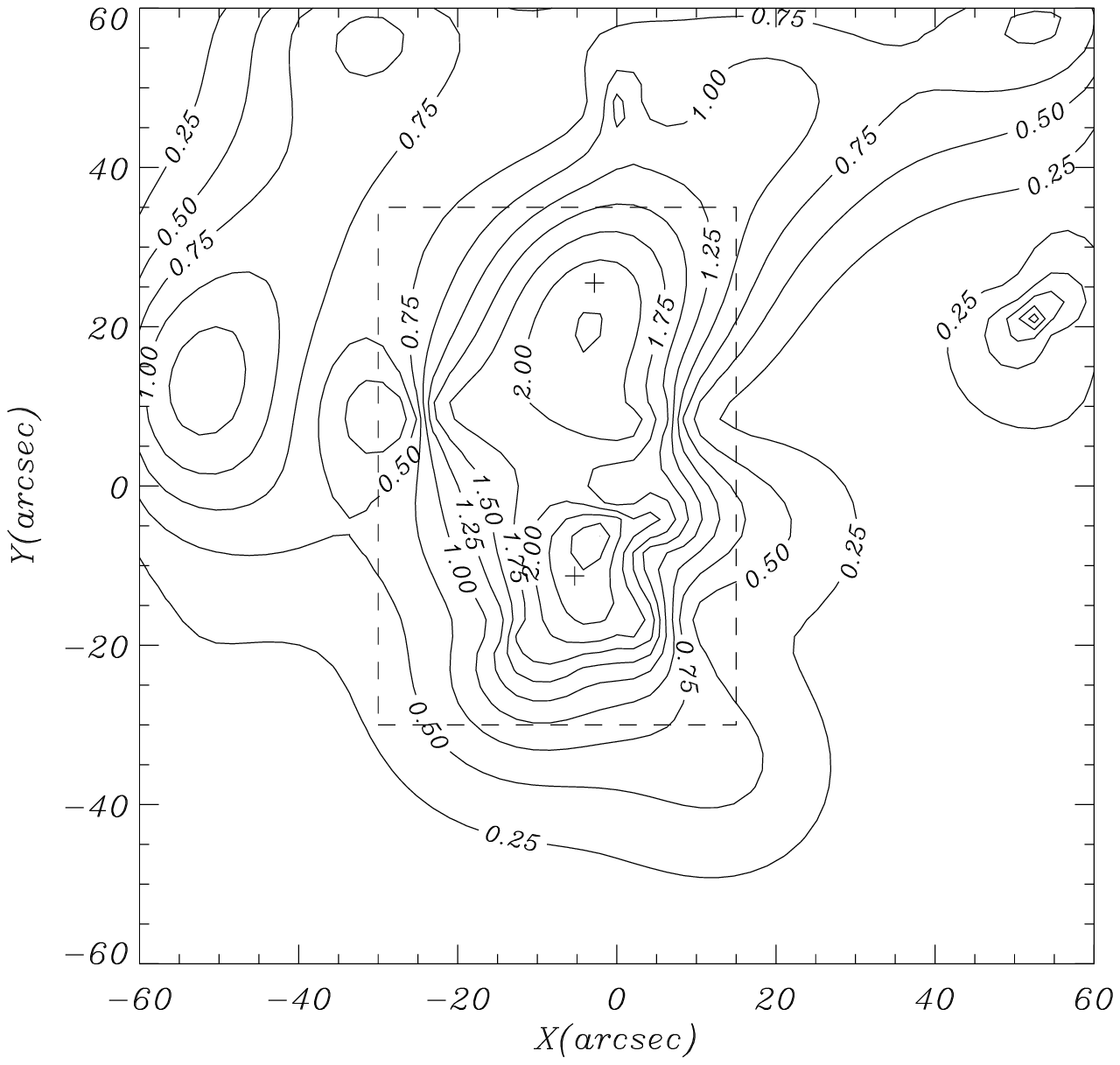}}
\parbox[c]{0.46\textwidth}{%
\epsfxsize=\hsize\epsffile{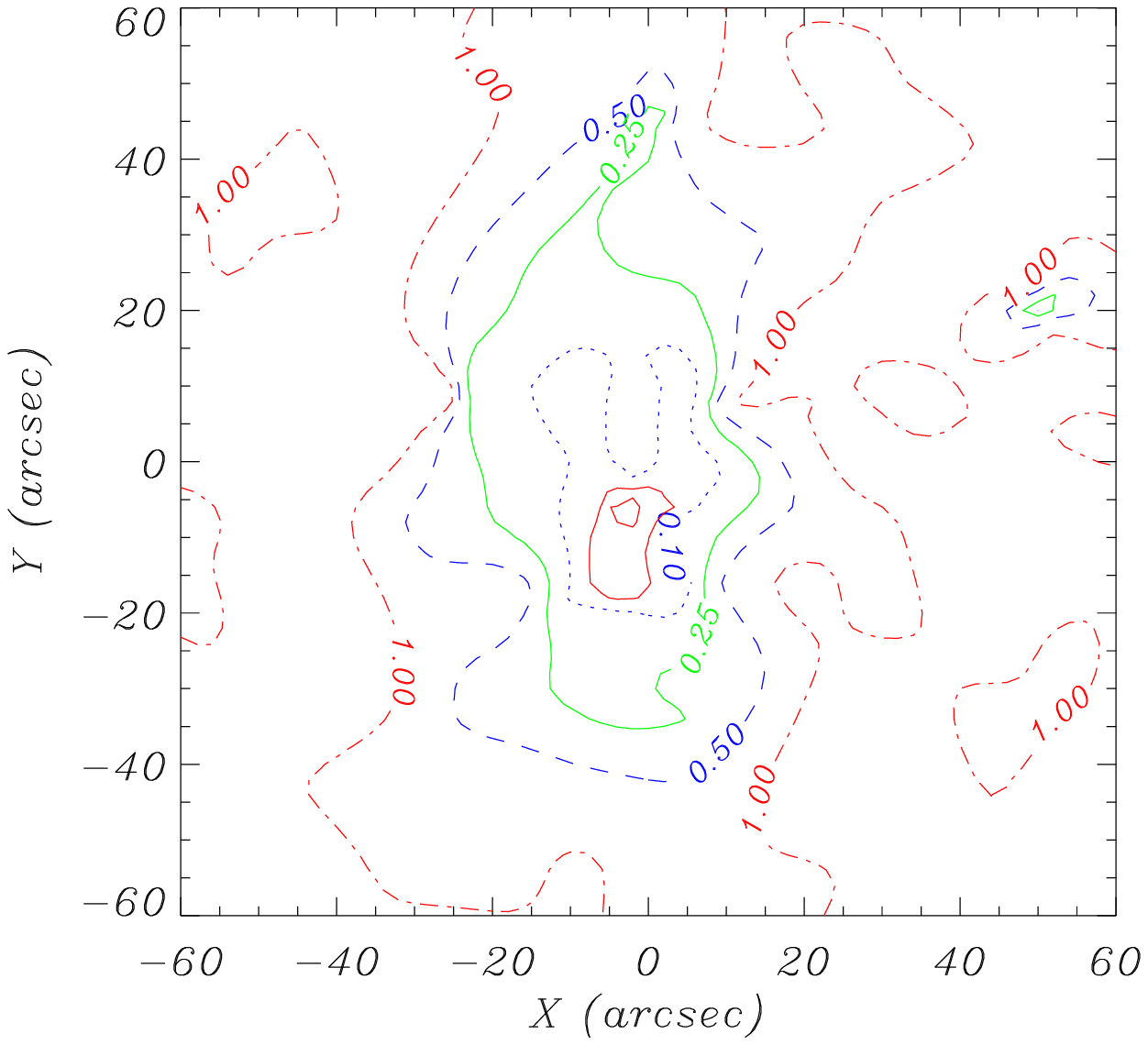}} 
\caption{\small {\bf Left:} Sky projected mass distribution
of Abell 370. Mass enclosed by the dashed rectangle is $2.1\times
10^{14}h^{-1}_{50}M_{\odot}$. {\bf Right} Fractional uncertainty of
the mass map.} 
\end{figure}
Our reconstruction of Abell 370, using only the constraints from
multiple images (strong lensing), is shown in Fig.~1. The reconstructed
mass map predicted the bimodal nature of the cluster and robustly
revealed features not associated with light (for more details see
AbdelSalam et al 1997). 

Our reconstruction indicates that the radial arc, recently discovered
in the HST image, may be five images out of a seven image
configuration. 
\section{Abell 2218}
Our mass reconstruction for A2218, using the constraints from combined
strong and weak lensing, is shown in Figure~2 (left) and the uncertainties
are shown in Figure~3. 
\begin{figure}[htb]
  \parbox[c]{0.50\textwidth}{%
    \epsfxsize=\hsize\epsffile{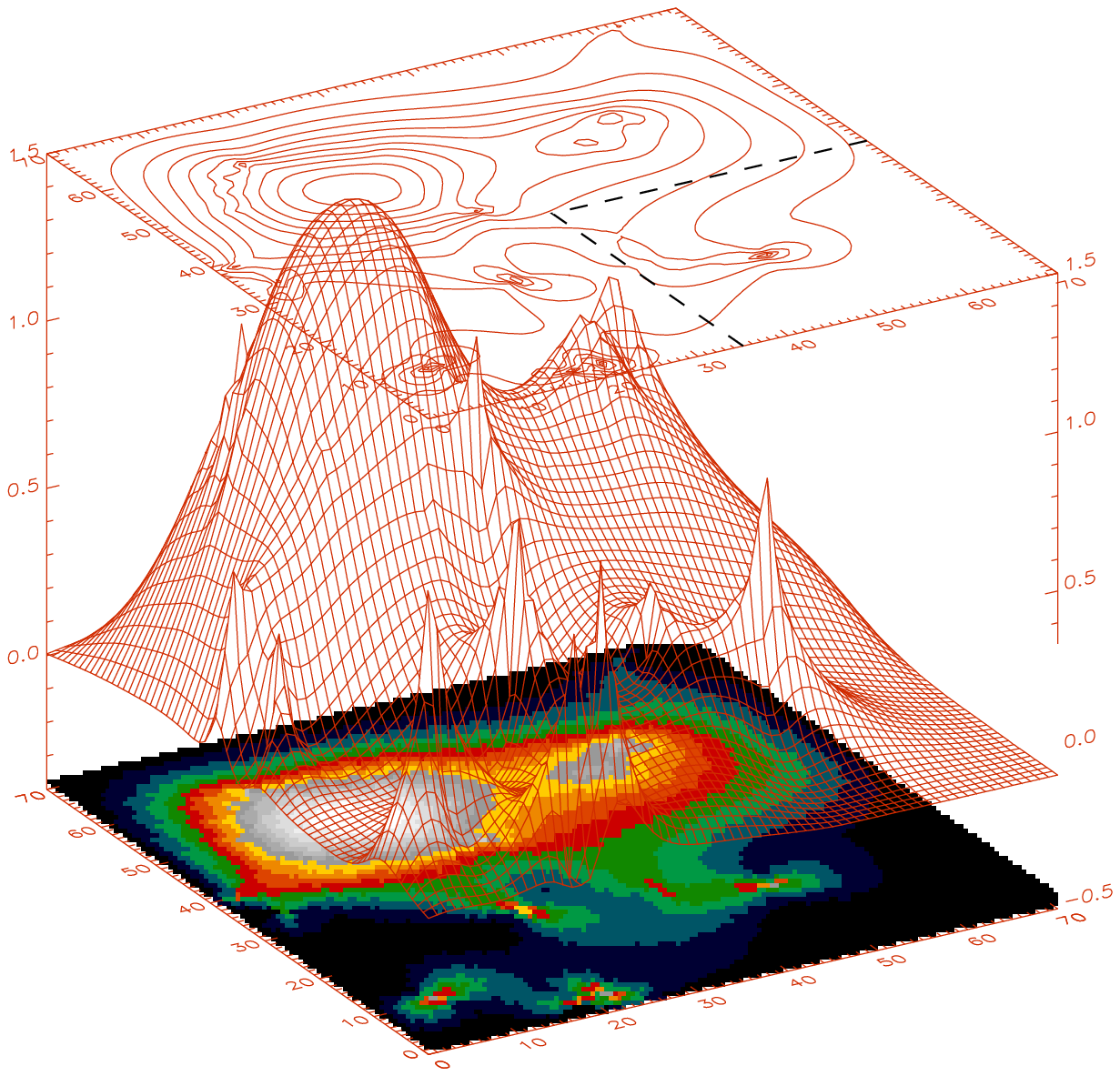}}
  \parbox[c]{0.45\textwidth}{%
    \epsfxsize=\hsize\epsffile{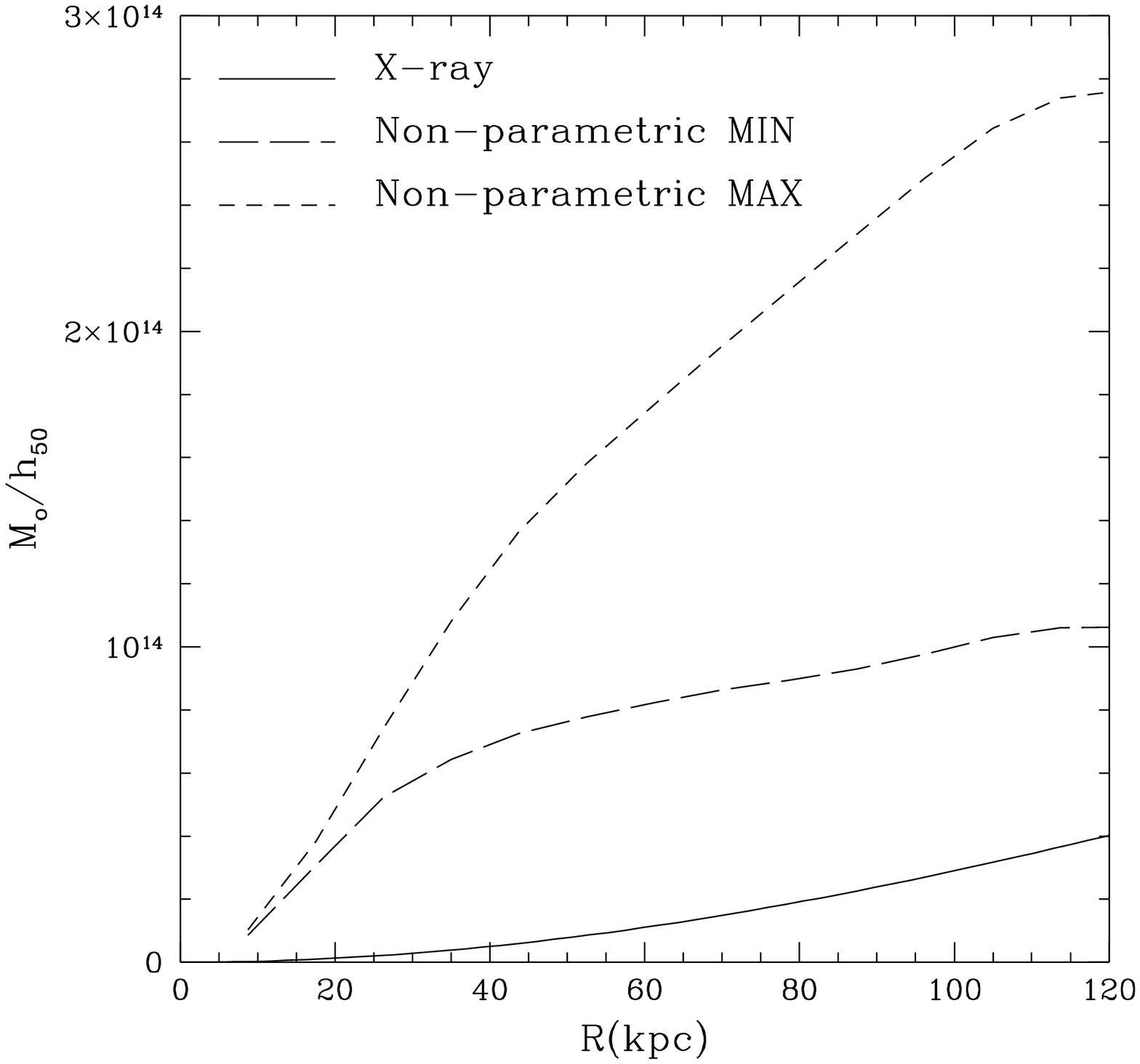}}
\caption{\small {\bf Left:} Mass distribution of Abell 2218 in units of the
critical surface density $3.1\times 10^{10}h_{50}^{-1}M_{\odot}{\rm
arcsec}^{-2}$. The total mass in the region of the HST image is
$2.74\times 10^{14}h_{50}^{-1}M_{\odot}$. {\bf Right:} The range of
radial mass profiles for Abell 2218 allowed by lensing, and the mass
radial profile from an X-ray model by Allen \etal 1997.}  
\end{figure}
Our reconstructions of the inner region of Abell 2218 uses both
multiple image data (strong lensing) and singly imaged arc(lets) (weak
lensing) from HST, plus spectroscopic redshitfs from Ebbels \etal
1997.
\begin{figure}[htb]
  \parbox[c]{0.47\textwidth}{%
    \epsfxsize=\hsize\epsffile{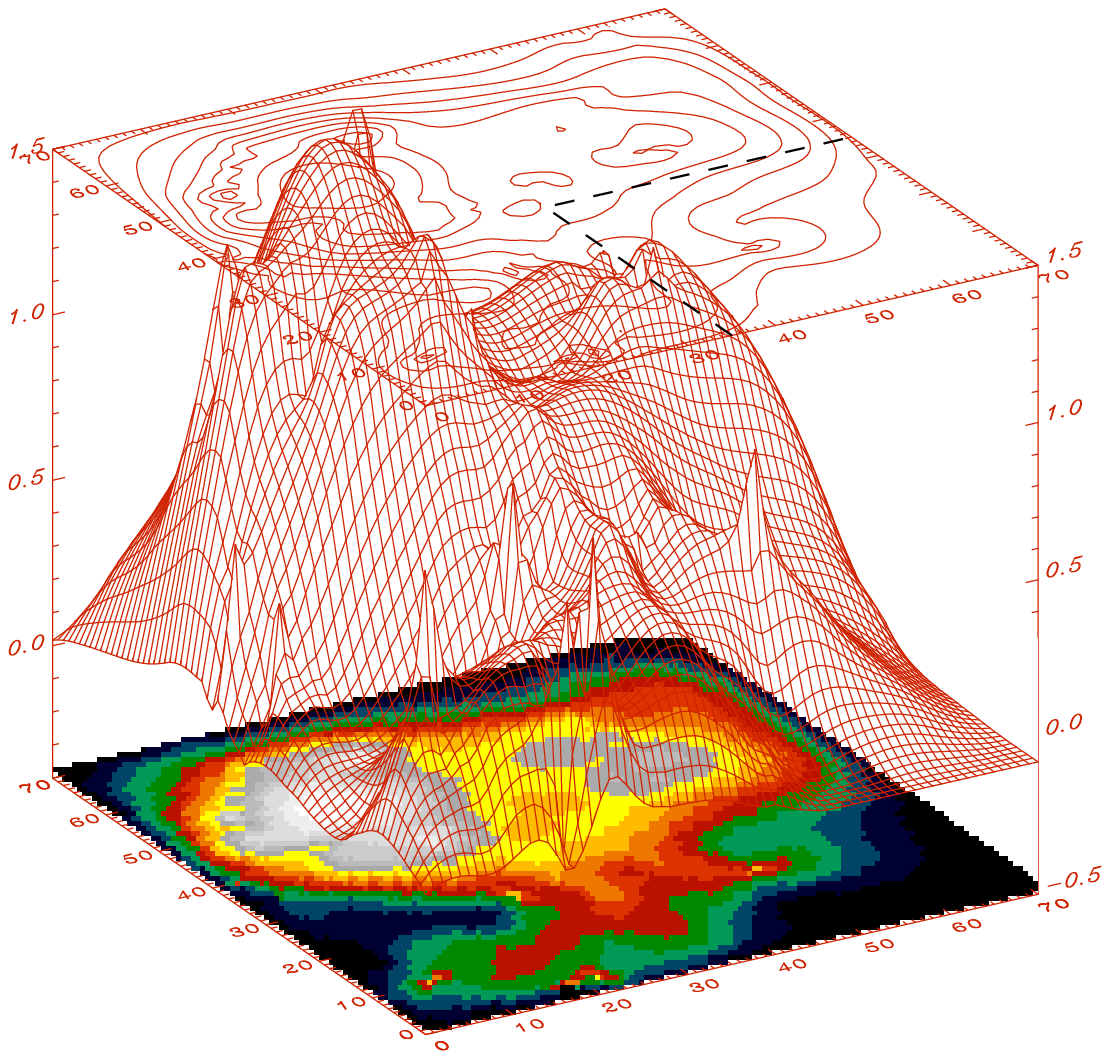}}
   \parbox[c]{0.47\textwidth}{%
     \epsfxsize=\hsize\epsffile{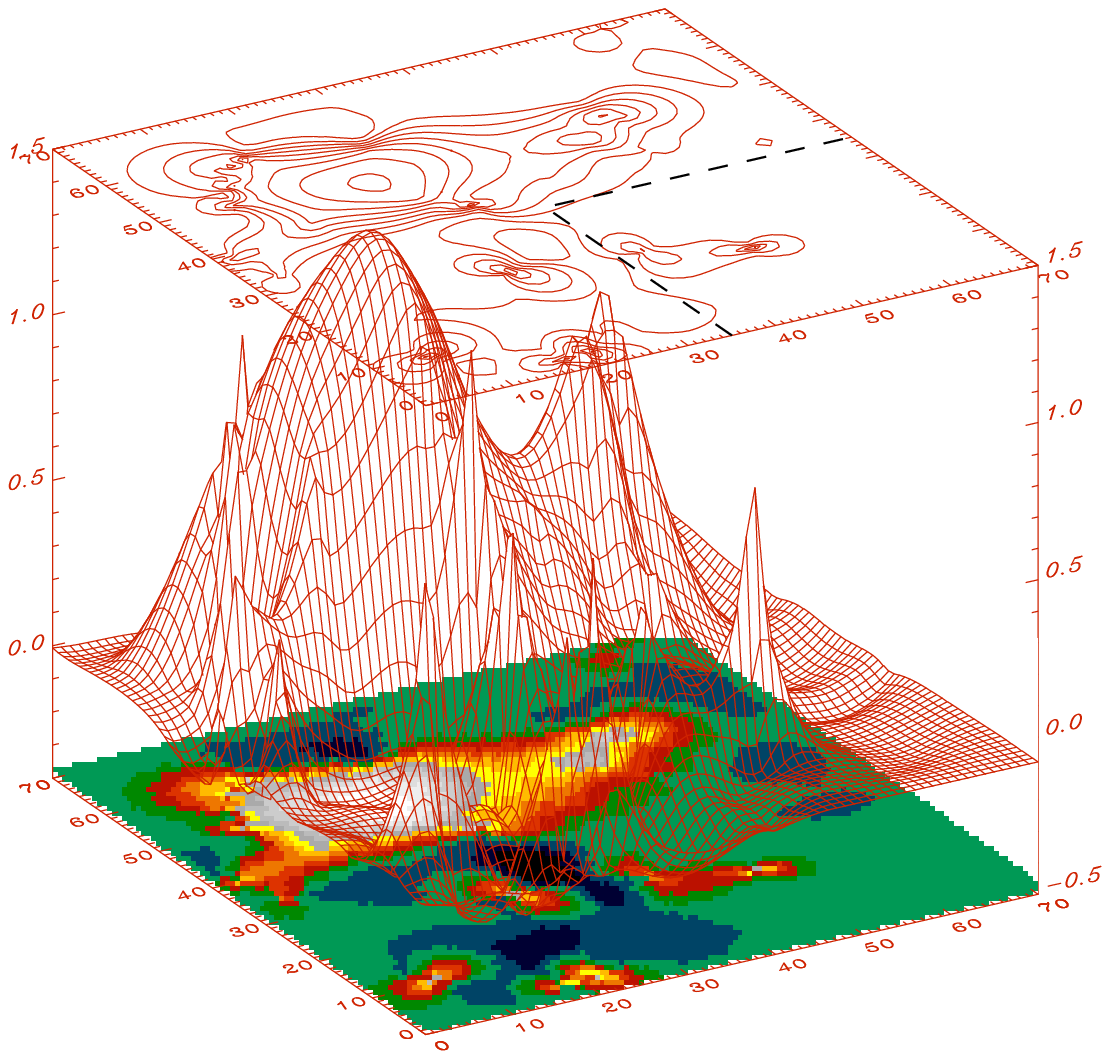}} 
\caption{\small {\bf Left:} Mass distribution
of Abell 2218 plus ({\bf left}) and minus ({\bf right}) one standard
deviation uncertainty.} 
\end{figure}
As far as we are aware this is the first mass reconstruction using
strong and weak lensing simultaneously.

As Miralda-Escud\'e \& Babul (1995) first noted, current X-ray
mass models appear to underestimate the mass. Our results reinforce
this conclusion: mass estimate from lensing is at least 2.5 times that
from X-ray models (Figure~2 (right)). This may indicate the gas is
partially supported by turbulence or magnetic fields, the gas is not
uniform and isothermal or is not in equilibrium, but rather multiphase
or has temperature gradient rising inwards; or the cluster is
elongated along the line of sight. 

\section*{Acknowledgements}
HMA acknowledgs the support of Overseas Research Scheme,
Oxford Overseas Bursary and Wolfson College (Oxford) Bursary. LLRW
acknowledgs the support of PPARC fellowship at the IoA, Cambridge. HMA
\& LLRW thank the organisers of {\it Large Scale Structure: Tracks and
Traces} for their kind hospitality.

\end{document}